\documentclass[twocolumn,showpacs,amsmath,amssymb,nofootinbib,superscriptaddress]{revtex4}

\usepackage{graphicx}
\usepackage{epsfig}
\usepackage{bm}
\usepackage{amsfonts}
\usepackage{color}
\usepackage{float}
\usepackage{amsmath}                  
\usepackage{dcolumn}
\usepackage{hyperref}
\usepackage{array}
\usepackage{lscape}

\usepackage{array}
\usepackage{booktabs}
\usepackage{multirow}

\providecommand{\U}[1]{\protect\rule{.1in}{.1in}}

\def\be{\begin{equation}}
\def\ee{\end{equation}}
\def\bea{\begin{eqnarray}}
\def\eea{\end{eqnarray}}
\def\eqi{\begin{equation}}
\def\eqf{\end{equation}}
\def\eqia{\begin{eqnarray}}
\def\eqfa{\end{eqnarray}}

\newcommand{\udt}[3]{#1^{#2}_{\phantom{#2}#3}}

\newcommand{\dut}[3]{#1_{#2}^{\phantom{#2}#3}}

\newcommand{\udut}[4]{#1^{#2\phantom{#3}#4}_{\phantom{#2}#3}}

\allowdisplaybreaks

\begin{document}

\title{Gravitational Waves in Modified Teleparallel Theories}

\author{Gabriel Farrugia}\email{gabriel.farrugia.11@um.edu.mt}
\affiliation{Department of Physics, University of Malta, Msida, MSD 2080, Malta}
\affiliation{Institute of Space Sciences and Astronomy, University of Malta, Msida, MSD 
2080, Malta}

\author{Jackson Levi Said}\email{jackson.said@um.edu.mt}
\affiliation{Department of Physics, University of Malta, Msida, MSD 2080, Malta}
\affiliation{Institute of Space Sciences and Astronomy, University of Malta, Msida, MSD 
2080, Malta}

\author{Viktor Gakis}\email{Vgakis@central.ntua.gr}
\affiliation{Department of Physics, National Technical University of Athens, Zografou 
Campus GR 157 73, Athens, Greece}

\author{Emmanuel N. Saridakis}
\email{Emmanuel\_{}Saridakis@baylor.edu}
\affiliation{Department of Physics, National Technical University of Athens, Zografou
Campus GR 157 73, Athens, Greece}
\affiliation{CASPER, Physics Department, Baylor University, Waco, TX 76798-7310, USA}


%%%%%%%%%%%%%%
\begin{abstract}
We investigate the gravitational waves and their properties in various modified 
teleparallel theories, such as $f(T), f(T,B)$ and $f(T,T_G)$ gravities. We perform the 
perturbation analysis both around a Minkowski background, as well as in the case where a 
cosmological constant is present, and for clarity we use both the metric and tetrad 
language. For $f(T)$ gravity we verify the result that no further polarization modes 
comparing to general relativity are present at first-order perturbation level, and we 
show that in order to see extra modes one should look at third-order perturbations. For 
non-trivial $f(T,B)$ gravity, by examining the geodesic deviation equations, we show 
that extra polarization models, namely the longitudinal and breathing modes, do appear at 
first-order perturbation level, and the reason for this behavior is the fact that 
although the first-order perturbation does not have any effect on $T$, it does affect the 
boundary term $B$. Finally, for $f(T,T_G)$ gravity  we show that at 
first-order perturbations the gravitational waves exhibit the same behavior to those of 
$f(T)$ gravity. Since different modified teleparallel theories exhibit different 
gravitational wave properties, the advancing gravitational-wave astronomy would help to 
alleviate the degeneracy not only between curvature and torsional modified gravity but 
also between different subclasses of modified teleparallel gravities.

\end{abstract}

\maketitle

\section{Introduction}

The discovery of the late-time accelerating expansion of the Universe and the study of 
galactic rotation curves has generated a lot of interest and investigation, particularly 
in the direction of dark energy and dark matter 
\cite{Rubin:1970zza,Riess:1998cb,Riess:2004nr}. Additionally, it has led to 
investigations into gravitational theories beyond general relativity (GR), with the most 
studied cases being modifications of the Einstein-Hilbert action, which is 
constructed from the Ricci scalar $R$. Amongst others one may have $f(R)$ gravity 
\cite{DeFelice:2010aj,Nojiri:2010wj}, theories with
inclusion of other scalar invariants (for instance, $f(R,G)$ gravity where $G$ is the 
Gauss-Bonnet term \cite{Nojiri:2005jg,Bamba:2009uf}, and more generally   
Lovelock gravity \cite{Lovelock:1971yv,Deruelle:1989fj}), theories with non-minimal 
curvature-matter couplings (e.g. $f(R,\mathcal{T})$ gravity, where $\mathcal{T}$ 
is the trace of the stress-energy tensor 
\cite{Harko:2011kv,Momeni:2011am,Alvarenga:2013syu}), or more radical modifications such 
as massive gravity \cite{deRham:2010kj}, Ho\v{r}ava-Lifshitz \cite{Horava:2009uw}, etc. 
The goal of all these endeavors is to consistently explain the aforementioned 
observational phenomena while also retaining GR as a particular limit 
\cite{Clifton:2011jh}.  

Recently, there has been a significant rise in interest in a specific class of theories 
originally investigated by Einstein and Cartan. By considering gravitation to be 
described by torsion rather than   curvature, gravitation can retain many of 
the features present in the original GR formalism \cite{Einstein:1928gr,Cartan:1922cr}. 
This is most commonly referred to as teleparallel gravity. Furthermore, the 
fundamental dynamical quantity of the theory is not the metric tensor but the more subtle, 
so 
called, tetrad field. In the simplest form of these theories the 
Lagrangian is just the torsion scalar $T$, constructed by contractions of the torsion 
tensor, and variation with respect to the tetrad gives rise to exactly the same equations 
with GR, that is why this theory was name Teleparallel Equivalent of General Relativity 
(TEGR) \cite{Moller:1961cl,Pellegrini:1963tf,Cho:1975dh,Hayashi:1979qx}. 
 The source of the above equivalence is a boundary quantity, $B$, which relates 
the two Lagrangians, namely the Ricci scalar of GR and the torsion scalar of TEGR:
\begin{equation}
\label{relation1}
R=-T+B,
\end{equation}
where $R$ is calculated with the regular Levi-Civita connection while $T$ is calculated 
with the Weitzenb\"{o}ck connection. 

Inspired by the gravitational modifications that are based on the curvature formulation 
of gravity, one can construct modified gravity theories starting from TEGR. The simplest 
such  modified teleparallel theory is the $f(T)$ gravity, in which one generalizes $T$ to 
a function $f(T)$ in the Lagrangian 
\cite{Bengochea:2008gz,Ferraro:2006jd} (see \cite{Cai:2015emx} 
for a review). One can immediately see that due to relation (\ref{relation1}) and in 
particular to the boundary term, $f(T)$ gravity is not   equivalent to $f(R)$ gravity, 
and thus it is a novel gravitational modification.
Additionally, the advantage of this  theory is that the equations of motion are of second 
order, in contrast to the fourth-order equations of $f(R)$ gravity. These features 
led to many investigations in various fields 
of cosmology in this theory 
\cite{Chen:2010va,Bamba:2010wb,Wu:2010mn,Cai:2011tc,
Zhang:2011qp,Rezazadeh:2015dza,Farrugia:2016qqe,Hohmann:2017jao,Basilakos:2018arq}. 
Furthermore, 
one may 
proceed in 
constructing other modifications and 
extensions of teleparallel theories, such as the $f(T,T_G)$ gravity, where $T_G$ 
is the teleparallel equivalent of $G$  
\cite{Kofinas:2014owa,Kofinas:2014daa,Chattopadhyay:2014xaa}, the $f(T,\mathcal{T})$ 
gravity, where $\mathcal{T}$ is the trace of the stress-energy tensor 
\cite{Harko:2014aja,Saez-Gomez:2016wxb}, or torsional gravities with higher-order 
derivatives \cite{Otalora:2016dxe}. Finally, one interesting class of torsional 
gravitational modification is the $f(T,B)$ gravity, in which one allows for the use of 
the 
boundary term $B$ in the Lagrangian 
\cite{Bahamonde:2015zma,Wright:2016ayu,Bahamonde:2016grb}.

On the other hand, gravitational wave (GW) observations have confirmed not only the 
existence of gravitational waves as the mediator of gravitational information 
\cite{Abbott:2016blz}, but have also set bounds on the polarization modes of these waves 
from known sources \cite{Abbott:2017oio}, as well as on their speed, which is equal to 
the 
light speed with great accuracy 
\cite{Lombriser:2015sxa,Lombriser:2016yzn,Sakstein:2017xjx,Ezquiaga:2017ekz,Baker:2017hug}
. These 
observations are very important for alternative theories of gravity, since in general one 
can obtain extra polarization modes or variant speed. Although there have been some works 
investigating gravitational waves in $f(T)$ gravity 
\cite{Bamba:2013ooa,Abedi:2017jqx,Cai:2018rzd}, the systematic study of gravitational 
waves in modified teleparallel gravities has not been performed.

In this work we are interested in looking at $f(T)$, $f(T,B)$ and $f(T,T_G)$ gravities in 
the realm of gravitational waves through detailed perturbations analysis.
Our goal is to determine whether various teleparallel gravities predict extra modes, and 
to investigate the strength of these modes in the scales where they arise. Although in GR 
there are two GW polarizations, namely the plus and cross 
polarizations, alternative and extended theories might yield more modes (as for instance 
in $f(R)$ gravity \cite{Capozziello:2011et}). As we will show, although in the  case 
of $f(T)$ gravity the polarization models  are identical 
to those of GR \cite{Bamba:2013ooa}, this is not the case when an arbitrary boundary 
contribution is included, as for instance in $f(T,B)$ gravity.  

The paper is organized as follows. In Section \ref{Themodel} we briefly review 
teleparallel gravity and its various modifications. In Section \ref{GWinfT} we perform an 
analysis of the gravitational waves in the case of $f(T)$ gravity, both in the metric and 
tetrad language, and for both zero and non-zero spin connection. In Section 
\ref{GwinfTB2} we investigate the gravitational waves in the case of $f(T,B)$ gravity, 
both around a Minkowski background, case which is 
obtained in the absence of a cosmological constant, but also in the case where the 
presence of a cosmological constant changes the background around which the perturbations 
are realized. In Section \ref{GWinTTG} we examine the gravitational waves in  $f(T,T_G)$. 
Finally, the work closes with a discussion and conclusion of results in section 
\ref{Conclusion}.

\section{Modified Teleparallel Theories of Gravity}
\label{Themodel}

In teleparallel theories of gravity the fundamental dynamical variable is the tetrad (or 
vierbein) 
$\udt{e}{a}{\mu}$, which relates the standard coordinate frame $\partial_\mu \equiv 
\frac{\partial}{
\partial x^\mu}$ with an orthonormal and non-coordinate frame ($e$-frame). In general, 
non-coordinate frames are anholonomic, a property that is attributed to the existence of 
non-inertial 
effects. The metric tensor $g_{\mu\nu}$ can be related to the tetrad through the Minkowski 
metric $\eta_{ab}$ by 
\begin{equation}
g_{\mu\nu} = \eta_{ab}\udt{e}{a}{\mu} \udt{e}{b}{\nu},
\label{metrictetrad}
\end{equation}
where the point dependence is suppressed for brevity. In the whole manuscript Greek 
indices 
refer to the spacetime coordinates, while Latin 
indices refer to the tangent-space ones. The inverse tetrad is denoted by 
$\dut{E}{a}
{\mu}$ for transparency, and one can show that  
\begin{equation}\label{eq:identity_tetrad-condition}
\udt{e}{a}{\mu} \dut{E}{a}{\nu} = \delta^\nu_\mu, \ \ \ \udt{e}{a}{\mu} \dut{E}{b}{\mu} = 
\delta^a_b.
\end{equation}
The connection used in the teleparallel theories of gravity is defined as a connection 
that has 
vanishing curvature. This connection is the so called Weitzenb\"{o}ck connection and the 
fact that 
it is torsionful makes the connection coefficients non-symmetric in the lower indices in 
contrast 
with the Levi-Civita connection where the indices are symmetric. The tetrad enables us to 
relate to 
each Lorentz spin connection $\udt{\omega}{a}{b\mu}$ the Weitzenb\"{o}ck connection via 
\cite{aldrovandi2012teleparallel}
\begin{equation}
\label{eq:weitzenbockdef}
\widehat{\Gamma}^{\rho}_{\nu\mu} \equiv \dut{E}{a}{\rho} \partial_\mu \udt{e}{a}{\nu} + 
\dut{E}{a}{\rho} \udt{\omega}{a}{b\mu}\udt{e}{b}{\nu}.
\end{equation}
The spin connection $\udt{\omega}{a}{b\mu}$ does not represent any additional 
gravitational degrees 
of freedom. If one switches over to the $e$-frame and applies the Weitzenbock covariant 
derivative 
to the basis vectors of the $e$-frame, assuming the so called Weitzenbock condition where 
$\udt{\omega}{a}{b\mu} = 0$ then the result will be zero. This phenomenon is called 
complete 
frame induced 
parallelism and in the physics literature is frequently called Teleparallelism or 
absolute 
parallelism \cite{Hayashi:1979qx}. The Riemann and Ricci tensors calculated with the 
Weitzenb\"{o}ck connection are identically zero, while     
the torsion tensor writes as
\begin{equation}
\label{eq:torsiontensordef}
\udt{T}{a}{\mu\nu} \equiv \widehat{\Gamma}^{a}_{\nu\mu} - \widehat{\Gamma}^{a}_{\mu\nu} 
= \partial_
\mu \udt{e}{a}{\nu} - \partial_\nu \udt{e}{a}{\mu}    + 
\udt{\omega}{a}{b\mu}\udt{e}{b}{\nu} - \udt{\omega}{a}{b\nu}\udt{e}{b}{\mu}.
\end{equation}
 Moreover, one can define the \textit{superpotential tensor} as
\begin{equation}\label{eq:superpotentialdef}
\dut{S}{a}{\mu\nu}\equiv 
\frac{1}{2}\left(\udt{K}{\mu\nu}{a}+\dut{e}{a}{\mu}\udt{T}{\alpha\nu}{\alpha}-\dut{e}{a}{
\nu}\udt{T}{\alpha\mu}{\alpha}\right),
\end{equation}
where $\udt{K}{\mu\nu}{a}$ is the \textit{contorsion tensor} identified as
\begin{equation}\label{eq:contorsiondef}
\udt{K}{\mu\nu}{a} \equiv \dfrac{1}{2} \left(\dut{T}{a}{\mu\nu} + \udt{T}{\nu\mu}{a} - 
\udt{T}{\mu\nu}{a}\right),
\end{equation}
which represents the difference between the Levi-Civita connection and the 
Weitzenb\"{o}ck 
connection. The Lagrangian of TEGR is the torsion scalar $T$, constructed by contractions 
of the torsion tensor, namely \cite{aldrovandi2012teleparallel}
\begin{equation}
\label{eq:torsionscalardef}
T \equiv \dut{S}{a}{\mu\nu}\udt{T}{a}{\mu\nu}=
\frac{1}{4} T^{\rho \mu \nu} T_{\rho \mu \nu} + \frac{1}{2}T^{\rho \mu \nu }T_{\nu
\mu\rho}
- T_{\rho \mu }^{\ \ \rho }T_{\ \ \ \nu }^{\nu \mu} .
\end{equation}
Therefore, the action of TEGR reads
\begin{equation}
\label{eq:teleparallel-action}
S = \dfrac{1}{16 \pi G} \int d^4x \: e \: T + \int d^4x \: e \: \mathcal{L}_m,
\end{equation}
where $e = \det\left(\udt{e}{a}{\mu}\right) = \sqrt{-g}$, with $g$ being the determinant 
of the metric tensor, and where $\mathcal{L}_m$ is the matter Lagrangian and $G$ the 
Newton's constant.

As we mentioned in the 
introduction, one can show that the Ricci scalar calculated with the Levi-Civita 
connection, and the torsion scalar calculated with the Weitzenb\"{o}ck 
connection, are related through  
\begin{equation}
R = -T - 2\nabla^\mu \udt{T}{\nu}{\mu\nu},
\end{equation} 
and thus  we can identify the boundary term $B \equiv - 2\nabla^\mu 
\udt{T}{\nu}{\mu\nu}$. Hence, one can immediately see that GR and TEGR will lead to 
exactly the same equations. However, this will not be the case if one uses $f(R)$ and 
$f(T)$ as the Lagrangian of the theory, which therefore correspond to different 
gravitational modifications.

A general class of modified teleparallel gravity would thus be composed by an arbitrary 
function of $T$ and $B$, leading to $f(T,B)$ 
gravity \cite{Bahamonde:2015zma}, characterized by the action
\begin{equation} \label{eq:general-action}
S = \dfrac{1}{16 \pi G} \int d^4x \: e \: f(T,B) + \int d^4x \: e \: \mathcal{L}_m.
\end{equation}
By varying the action with respect to the vierbein we obtain the following field 
equations  
\begin{align}\label{eq:general-field-equations}
&\dut{E}{a}{\mu} \Box f_B - \dut{E}{a}{\nu} \nabla^\mu \nabla_\nu f_B + \dfrac{1}{2}B f_B 
\dut{E}{a}
{\mu} \nonumber \\
&+ 2\partial_\nu \left(f_B+f_T\right)\dut{S}{a}{\nu\mu} + 2e^{-1}\partial_\nu 
\left(e\dut{S}{a}{\nu\mu}\right)f_T \nonumber \\
&- 2f_T \udt{T}{\alpha}{\nu a}\dut{S}{\alpha}{\mu\nu} - \dfrac{1}{2}\dut{E}{a}{\mu} f = 
8\pi G\, \dut{\Theta}
{a}{\mu},
\end{align}
where $\dut{\Theta}{a}{\mu}$ is the stress-energy tensor, which in terms of the matter 
Lagrangian 
is given by $\dut{\Theta}{a}{\mu} = -\delta \mathcal{L}_\text{m}/\delta \udt{e}{a}{\mu}$. 
In the above equation we have defined that $f_{T}\equiv\partial f/\partial T$ and
$f_{B}\equiv\partial f/\partial B$. Note that the derived equations are given for the zero 
spin 
connection case. Additionally, in terms of spacetime indices the 
equations of motion can take the form
\begin{align}\label{eq:global-eq-motion}
&-f_T G_{\mu\nu} + \left(g_{\mu\nu} \Box - \nabla_\mu \nabla_\nu\right) f_B \nonumber \\
& + \dfrac{1}{2}g_{\mu\nu}\left(f_B B + f_T T - f\right) \nonumber \\
& + 2{{S_{\nu}}^\alpha}_\mu \partial_\alpha \left(f_T + f_B\right) = 8\pi 
G\,\Theta_{\mu\nu},
\end{align}
where $G_{\mu\nu} \equiv R_{\mu\nu} - \dfrac{1}{2}g_{\mu\nu}R$ is the Einstein tensor 
calculated with the Levi-Civita connection. 

Before closing this section let us make some comments on the spin connection 
$\udt{\omega}{a}{b\mu}$ that is present in the definition
(\ref{eq:weitzenbockdef}). In the traditional works of TEGR  one usually 
sets it to zero for convenience, by choosing a suitable frame (specifically 
autoparallel orthonormal frame) \cite{aldrovandi2012teleparallel}. Although this does 
not have any effect for TEGR, in the case of $f(T)$ gravity such a preferred frame 
choice should be used carefully. In particular, one is allowed to  make such a choice in 
order to find cosmological solutions, however one has to have in mind that in 
investigations which include questions on Lorentz transformations such a formulation is 
in general inadequate. In this case one should formulate $f(T)$ gravity in a fully 
covariant way, keeping a general non-zero spin connection \cite{Krssak:2015oua}. In this 
way the theory becomes completely consistent with  Lorentz invariance, nevertheless at 
the price of increased complication. In the largest part of this manuscript we will 
consider the zero spin connection case (which is a safe choice for 
cosmological applications), especially for the general $f(T,B)$ and $f(T,T_G)$ theories,  
however in the simpler $f(T)$ theory, for completeness, we will also discuss the non-zero 
spin connection case.

\section{Gravitational Waves in $f(T)$ gravity}
\label{GWinfT}

Let us now start with the investigation of the gravitational waves in the case of the 
simplest modification to teleparallel gravity, namely $f(T)$ gravity. We first perform 
the analysis for first order perturbations at a tetrad level, and then we proceed to  
higher-order examination in order to understand  more 
transparently the potential deviations from GR.

\subsection{Tetrad Solutions for GWs in $f(T)$ gravity} 
\label{Tetrad_Solutions_for_GWs}

%In the previous analysis we studied the gravitational waves from the metric 
%perturbation side. For completeness, and for clarity, we now proceed to their 
%examination 
%from the tetrad perturbation side.

We start by considering the tetrad form of the field  equations 
(\ref{eq:global-eq-motion}) in the case of $f(T)$ gravity, namely  
\begin{align}
&e^{-1}f_T \partial_\nu \left(e \dut{S}{a}{\mu\nu}\right) + f_{TT} \dut{S}{a}{\mu\nu} 
\partial_\nu 
T \nonumber\\
&- f_T \udt{T}{b}{\nu a}\dut{S}{b}{\nu\mu} + \dfrac{1}{4}f(T) \udt{e}{a}{\mu} = 0,
\label{eqfTtetrad}
\end{align}
where as before we neglect the matter sector. We consider a tetrad 
perturbation  of the form
\begin{equation}
\label{tetradpert1}
\udt{e}{a}{\mu} = \gamma^{(0)a}_\mu + \gamma^{(1)a}_\mu + 
\mathcal{O}\left(\gamma^{(2)a}_\mu\right),
\end{equation}
where $|\gamma^{(i)a}_\mu| << 1$ except for the zeroth order contribution, and each 
successive 
order is much smaller than the preceding order, i.e. 
$|\gamma^{(2)a}_\mu|<<|\gamma^{(1)a}_\mu|$. 
This last comment applies throughout to every successive order quantity. Throughout the 
work, 
superscripts with parenthesis will represent the perturbative order of the quantity being 
presented.
 As usual, from (\ref{metrictetrad}) we obtain for the 
zeroth order perturbation:
\begin{equation}\label{eq:background-Mink-condition}
\eta_{\mu\nu} = \eta_{ab} \gamma^{(0)a}_\mu \gamma^{(0)b}_\nu.
\end{equation} 
Thus, the torsion tensor (\ref{eq:torsiontensordef}),  assuming for the moment zero 
spin connection, 
up to first order can be expressed as
\begin{align}
\udt{T}{a}{\mu\nu} = \partial_\mu \gamma^{(0)a}_\nu - \partial_\nu \gamma^{(0)a}_\mu   
  + \partial_\mu \gamma^{(1)a}_\nu - 
\partial_\nu \gamma^{(
1)a}_\mu  .
\end{align}  
This gives an impression that an arbitrary choice of $\gamma^{(0)a}_\nu$ will yield a 
zeroth order 
contribution. However, as discussed in Ref.\cite{Krssak:2015oua},  if the gravitational 
strength, i.e. the gravitational constant, vanishes, one obtains the Minkowski 
background, in which the 
torsion tensor vanishes.   In this 
perturbation regime $\udt{e}{a}{\mu}|_{G \rightarrow 0} = 
\gamma^{(0)a}_\mu$, since this corresponds 
to the Minkowski 
background, while the higher-order perturbations are due to gravitational effects. 
Therefore
\begin{align}
\udt{T}{a}{\mu\nu}|_{G \rightarrow 0} = \partial_\mu \gamma^{(0)a}_\nu - \partial_\nu 
\gamma^{(0)a}_
\mu  = 0.
\end{align} 
Hence, as we mentioned above, the torsion tensor is first-order, and so are the 
contorsion 
and superpotential, which 
ultimately imply that the torsion scalar is second-order at the level of perturbations. 
Finally, in order to handle the $f(T)$ term we will consider the Taylor 
expansion (\ref{TaylorfT1}).

Inserting the above expressions into Eq. (\ref{eqfTtetrad}), order by order leads to
\begin{align}
 \label{eqaux200}
&\gamma^{(0)\rho}_a f^{(0)} = 0, \\
&f_T^{(0)}
e^{-1(0)} \partial_\nu \left(e^{(0)} \dut{S}{a}{(1)\mu\nu} \right) 
 + \gamma^{(0)\rho}_a \dfrac{f^{(1)}}{4} + \gamma^{(1)\rho}_a \dfrac{f^{(0)}}{4} = 0.
 \label{eqaux2}
\end{align}
As mentioned above, the first condition implies that no cosmological constant is 
present. We next identify 
that $e^{(0)}
 = e^{-1(0)} = 1$, and we assume   $f^{(1)} = 0$, since $T$ is a 
second-order quantity and thus its function cannot have first-order contributions. 
Lastly, we focus on the non-trivial case $f_T^{(0)}\neq0$. Under these considerations Eq. 
(\ref{eqaux2}) becomes  
\begin{align}
\partial_\nu \dut{S}{a}{(1)\mu\nu} = 0.
\label{eqcase1}
\end{align}
We note here that these intermediate steps are different from the $f(T)$ GWs analysis 
carried out previously with the Einstein tensor, since we now follow the tetrad language. 
Moreover, notice that this equation appears in TEGR too
\cite{Obukhov:2002hy}.

Let us proceed by extracting explicit solutions. We first remark however that the above 
equation is 
not possible to solve in general. However, we can assume that GR
gauge conditions on the perturbed metric, namely $h^{(1)}_{\mu\nu}$, can also be imposed 
here too, specifically that it is 
traceless 
\begin{equation}
\udt{h}{(1)\mu}{\mu} = 2\eta^{\mu\nu}\eta_{ab}\gamma^{(0)a}_\mu \gamma^{(1)b}_\nu = 0,
\end{equation}
and satisfies the Lorenz gauge condition
\begin{align}
&0 = \partial^\mu h^{(1)}_{\mu\nu} = \partial_b \gamma^{(0)b}_\nu \nonumber \\
&+ \eta_{ab} \left[\gamma^{(1)b}_\nu \partial^\mu \gamma^{(0)b}_\nu + \gamma^{(1)a}_\mu 
\partial^\mu \gamma^{(0)a}_\mu + \gamma^{(1)a}_\mu \partial^\mu \gamma^{(0)b}_\nu\right].
\end{align}
Together with the relation
\begin{equation}
\gamma^{(0)d}_\nu = \eta^{cd} \eta_{\mu\nu} \gamma^{(0)\mu}_c,
\end{equation}
and for simplicity we consider the 
case $\gamma^{(0)a}_\mu = \delta^a_\mu$, equation (\ref{eqcase1}) can be 
solely 
expressed in terms 
of $\gamma^{(1)a}_\mu$, namely
\begin{align}
&A^\mu_d \equiv \eta^{\mu\alpha} \eta_{df} \Box \gamma^{(1)f}_\alpha + \delta^\mu_a 
\delta^\rho_d \Box  \gamma^{(1)a}_\rho = 0.
\end{align}
This yields the following system of equations
\begin{align}
&A^0_0: & & \Box \gamma^{(1)0}_0 = 0, \label{eq:no-spin_eq1} \\
&A^0_i = -A_0^i: & &\Box \left(\gamma^{(1)0}_i - \gamma^{(1)i}_0\right) = 0, \\
&A^i_j \left(i \neq j\right): & &\Box \left(\gamma^{(1)j}_i + \gamma^{(1)i}_j\right) = 0, 
\\
&A^i_m \left(i = m\right): & &\Box \gamma^{(1)i}_i = 0, \label{eq:no-spin_eq4}
\end{align}
where we have used the fact that $\eta_{\mu\nu} = \text{diag}\left(-1,1,1,1\right)$ and 
$i, j = \{1,
2,3\}$. Since we are working in the Minkowski metric Cartesian coordinate system, the 
indices $\{0,
1,2,3\}$ correspond to $\{t,x,y,z\}$ respectively. Lastly, we can demand the extra gauge 
condition that the waves are transverse, i.e. $h^{(1)}_{0\mu} = 0$, which sets 
$\gamma^{(1)0}_0 = 0$ and $\gamma^{(1)i}_0 = \gamma^{(1)0}_i$. In summary, the full list 
of conditions and  equations are found 
to be
\begin{align}
& \text{Traceless condition:} &  &\gamma^{(1)i}_i = 0, \\
& \text{Lorenz gauge condition:} &  &\partial_j\left(\gamma^{(1)j}_i + \gamma^{(1)i}_j 
\right) = 0, 
\\
&A^i_j \left(i \neq j\right): & &\Box \left(\gamma^{(1)j}_i + \gamma^{(1)i}_j\right) = 0, 
\\
&A^i_m \left(i = m\right): & &\Box \gamma^{(1)i}_i = 0.
\end{align}

Without loss of generality we make the choice that the gravitational wave   propagates
in the 
$z$-direction, and as usual we work in the Fourier space. The wave equations then imply 
that
\begin{align}
 \gamma^{(1)i}_i &= {A_i}^i \exp (ik_\mu x^\mu), \nonumber\\
 i &= \{1,2,3\} \; \text{(fixed index)}, \\
 \nonumber\\
 \gamma^{(1)j}_i + \gamma^{(1)i}_j &= {B_i}^j \exp (ip_\mu x^\mu), \nonumber\\
 i &= \{1,2,3\}, \; i \neq j,
\end{align}
where $k_\mu$ and $p_\mu$ are wave-vectors such that $k_\mu k^\mu = p_\mu p^\mu = 0$, and 
where 
${A_i}^i$ and 
${B_i}^j$ are coefficients such that ${A_1}^1 = -{A_2}^2$, ${A_3}^3 = 0$ and ${B_1}^3 = 
{B_2}^3 = 
0$. Note that these conditions arise from the traceless and Lorenz gauge conditions. 
Therefore, the 
undetermined coefficients are ${A_1}^1, {B_1}^3, {B_2}^3$ and ${B_1}^2$, which leads to 
the 
perturbed tetrad solution
\begin{equation}
\gamma^{(1)a}_\mu =  \begin{pmatrix}
0 & \gamma^{(1)1}_0 & \gamma^{(1)2}_0 & \gamma^{(1)3}_0 \\
\gamma^{(1)1}_0 & \gamma^{(1)1}_1 & \gamma^{(1)2}_1 & \gamma^{(1)3}_1 \\
\gamma^{(1)2}_0 & {B_1}^2 \exp (ip_\mu x^\mu)-\gamma^{(1)2}_1 & -\gamma^{(1)1}_1 & 
\gamma^{(1)3}_2 \\
\gamma^{(1)3}_0 & -\gamma^{(1)3}_1 & -\gamma^{(1)3}_2 & 0
\end{pmatrix}.
\label{tetradsolbasic1}
\end{equation}
Here the $\gamma^{(1)j}_i$ are undetermined tetrad components which are not 
constrained by the equations.  We mention that the perturbed metric then takes the form
\begin{equation}
h^{(1)}_{\mu\nu} =   \begin{pmatrix}
0 & 0 & 0 & 0 \\
0 & 2\gamma^{(1)1}_1 & {B_1}^2 \exp (ip_\mu x^\mu) & 0 \\
0 & {B_1}^2 \exp (ip_\mu x^\mu) & -2\gamma^{(1)1}_1 & 0 \\
0 & 0 & 0 & 0
\end{pmatrix}.
\label{solutiongen1}
\end{equation}
Note that   obtaining the perturbed tetrad is  not a trivial task in general, since 
amongst the infinite choices of perturbed tetrad ansatzes corresponding to the same 
perturbed metric, one should use the 
appropriate ones in order to obtain consistency  
\cite{Chen:2010va,Zheng:2010am,Cai:2018rzd}.

Observing the solution (\ref{solutiongen1}) we can easily identify the standard $+$ and 
$\times$ polarizations of GR by 
defining $h_+ \equiv 2\gamma^{(1)1}_1$ and $h_\times \equiv {B_1}^2 \exp (ip_\mu x^\mu)$. 
Therefore, the perturbed 
tetrad has two physical degrees of freedom   ($h_+$ and $h_\times$) and 
six arbitrary degrees of freedom related to Lorentz transformations. Hence, through the 
explicit solutions we did verify the result obtained previously, namely that at 
first-order perturbation level there are not any new polarization modes in $f(T)$ 
gravity. Finally, as we mentioned above, in order to examine the six degrees of freedom 
related to Lorentz transformations it is necessary to re-formulate the theory in a fully 
covariant way, namely keeping an arbitrary spin connection   $\udt{\omega}{a}{b\mu}$. 
This is performed in the next subsection.

\subsection{GWs in $f(T)$ gravity with non-zero spin connection} 
\label{solunonzerospinconnection}

For completeness and transparency, in this subsection we perform the analysis of the 
previous subsection, but in the case of a general  spin connection, i.e. for the  
fully covariant formulation of $f(T)$ gravity presented in \cite{Krssak:2015oua}.

We start by considering the tetrad perturbation (\ref{tetradpert1}), however we  
insert it in the   torsion tensor (\ref{eq:torsiontensordef}) maintaining an
arbitrary spin connection,  obtaining   up to first order: 
\begin{eqnarray}
&&
\!\!\!\!\!\!\!\!\!\!\!\!\!\!
\udt{T}{a}{\mu\nu} = \partial_\mu \gamma^{(0)a}_\nu - \partial_\nu \gamma^{(0)a}_\mu + 
\udt{\omega}
{a}{b\mu}\gamma^{(0)b}_\nu - \udt{\omega}{a}{b\nu}\gamma^{(0)b}_\mu\nonumber \\
&&\ \ \   + \partial_\mu \gamma^{(1)a}_\nu - 
\partial_\nu \gamma^{(
1)a}_\mu   + \udt{\omega}{a}{b\mu}\gamma^{(1)b}_\nu - 
\udt{\omega}{a}{b\nu}\gamma^{(1)b}_\mu.\
\end{eqnarray}
As discussed in  \cite{Krssak:2015oua}, the purely inertial spin connection can be 
found by 
demanding that the torsion tensor is zero when the gravitational constant vanishes, 
namely $G 
\rightarrow 
0$, which yields the expression
\begin{equation}
\udt{\omega}{a}{b\mu} = \Gamma^a_{b\mu} - \dut{E}{b}{\nu} \partial_\mu 
\udt{e}{a}{\nu}|_{G\rightarrow 0},
\end{equation}
where $\Gamma^a_{b\mu}$ is the GR Levi-Civita connection. As before, $\udt{e}{a}{\mu}|_{G 
\rightarrow 0} = \gamma^{(0)a}_\mu$. 
Furthermore, 
$g_{\mu\nu}|_{G \rightarrow 0} = \eta_{\mu\nu}$ and hence $\Gamma^a_{b\mu}|_{G 
\rightarrow 0} = 0$. 
Therefore, the spin connection turns out to be
\begin{equation}
\udt{\omega}{a}{b\mu} = - \gamma^{(0)\nu}_b \partial_\mu \gamma^{(0)a}_\nu.
\end{equation}
Since the torsion tensor is zero when the gravitational constant is zero, then
\begin{equation}
\udt{T}{a}{\mu\nu}|_{G \rightarrow 0}= \partial_\mu \gamma^{(0)a}_\nu - \partial_\nu 
\gamma^{(0)a}
_\mu + \udt{\omega}{a}{b\mu}\gamma^{(0)b}_\nu   - \udt{\omega}{a}{b\nu}\gamma^{(0)b}_\mu 
= 
0.
\end{equation}
Thus, the torsion tensor is first-order, and so are the contorsion and superpotential,
which ultimately implies that the torsion scalar is second-order at the level of 
perturbations. 

Before 
investigating the field equations we make the following remark. The purely 
inertial spin 
connection is   given by
\begin{equation}
\udt{\omega}{a}{b\mu} = -\dut{\Lambda}{b}{c} \partial_\mu \udt{\Lambda}{a}{c},
\end{equation}
where $\dut{\Lambda}{d}{c}$ is a Lorentz matrix with inverse $\udt{\Lambda}{a}{c}$. Thus, 
under this formulation, we deduce that the zeroth-order tetrad perturbations 
$\gamma^{(0)a}_\nu$ 
are 
precisely the Lorentz matrices. This shall be considered in what follows. 

The next step 
is to 
expand the field equations at a perturbation level. In this case, the field equations for 
$f(T)$ 
gravity with an arbitrary spin connection are given by
\begin{align}
e^{-1}f_T &\partial_\nu \left(e \dut{S}{a}{\mu\nu}\right) + f_{TT} \dut{S}{a}{\mu\nu} 
\partial_\nu 
T - f_T \udt{T}{b}{\nu a}\dut{S}{b}{\nu\mu} \nonumber\\
& + f_T \udt{\omega}{b}{a\nu} \dut{S}{b}{\nu\mu} + \dfrac{1}{4}f(T) \udt{e}{a}{\mu} = 0,
\label{eqfTtetrad-spin}
\end{align}
which under the Taylor expansion (\ref{TaylorfT1}), expanding order by order we 
obtain  
\begin{align}
&\gamma^{(0)\rho}_a f^{(0)} = 0, \\
&f_T^{(0)}\left[e^{-1(0)} \partial_\nu \left(e^{(0)} \dut{S}{a}{(1)\mu\nu} \right) + 
\udt{\omega}{b}
{a\nu}\dut{S}{b}{(1)\nu\mu}\right] \nonumber\\
& + \gamma^{(0)\rho}_a \dfrac{f^{(1)}}{4} + \gamma^{(1)\rho}_a \dfrac{f^{(0)}}{4} = 0,
\label{eqaux2bb}
\end{align}
which generalize (\ref{eqaux200}),(\ref{eqaux2}) in the case of non-zero spin connection.
As before, the first condition implies that no cosmological constant is present. 
Moreover, we choose  $e^{(0)}
 = e^{-1(0)} = 1$, since $\det \gamma^{(0)a}_\nu = \det \udt{\Lambda}{a}{\nu} = 1$, which 
is a property of Lorentz matrices, and similarly to the zero spin connection case we 
impose  $f^{(1)} = 0$ and we assume $f_T^{(0)}\neq0$. Hence,  
Eq. 
(\ref{eqaux2bb}) becomes  
\begin{align}
\partial_\nu \dut{S}{a}{(1)\mu\nu} + \udt{\omega}{b}{a\nu}\dut{S}{b}{(1)\nu\mu} = 0,
\label{eqaux2cc}
\end{align}
which is the generalization of (\ref{eqcase1}) in the case of non-zero spin connection.
 Finally, using the definition of the spin connection $\udt{\omega}{a}{b\mu} = - 
\gamma^{(0)\nu}_b 
\partial_\mu \gamma^{
(0)a}_\nu$, equation (\ref{eqaux2cc}) can be recast into the simpler form
\begin{align}
&\gamma^{(0)\alpha}_a \partial_\nu \left[\gamma^{(0)b}_\alpha 
\dut{S}{b}{(1)\mu\nu}\right] \nonumber \\
&= \partial_\nu \dut{S}{a}{(1)\mu\nu} + \gamma^{(0)\alpha}_a \partial_\nu 
\gamma^{(0)b}_\alpha \dut{
S}{b}{(1)\mu\nu} = 0.
\end{align}

Similar to the spin zero case, solving the above equation  is not possible in general. 
However, we 
will again assume that the GR
gauge conditions on the perturbed metric being traceless and satisfies the Lorenz gauge 
condition 
can also be imposed 
here too. Hence, together with the relation
\begin{equation}
\gamma^{(0)d}_\nu = \eta^{cd} \eta_{\mu\nu} \gamma^{(0)\mu}_c,
\end{equation}
the equation of motion reduces to the following simplified expression
\begin{align}
&\eta^{\mu\alpha} \eta_{\beta\rho} \gamma^{(0)\beta}_d \Box\left(\gamma^{(0)\rho}_f 
\gamma^{(1)f}_\alpha\right) - \gamma^{(0)\beta}_d \partial_b \gamma^{(0)\mu}_c \partial^c 
\gamma^{(1)b}_\beta \nonumber \\
&+ \gamma^{(0)\beta}_d \Box\left(\gamma^{(0)\mu}_b \gamma^{(1)b}_\beta\right) - 
\gamma^{(0)\beta}_d 
\partial_\nu \gamma^{(1)c}_\beta \partial^\mu \gamma^{(0)\nu}_c \nonumber \\
&- \eta^{\mu\alpha} \partial_b \gamma^{(0)\beta}_d \partial_\beta \gamma^{(1)b}_\alpha + 
\eta^{\mu\alpha} \partial_d \gamma^{(0)\beta}_b \partial_\beta \gamma^{(1)b}_\alpha = 0,
\end{align}
which can alternatively   be expressed in terms of the spin connection as
\begin{align}
&\eta^{\mu\alpha} \eta_{df} \Box \gamma^{(1)f}_\alpha + \gamma^{(0)\beta}_d 
\gamma^{(0)\mu}_b \Box \gamma^{(1)b}_\beta + 2 \eta^{\mu\alpha} \omega_{df\nu} 
\partial^\nu \gamma^{(0)f}_\alpha \nonumber \\
&+ \eta^{\mu\alpha} \gamma^{(0)\beta}_d \gamma^{(1)f}_\alpha \partial_\nu 
\dut{\omega}{\beta f}{\nu}
 - \gamma^{(0)\beta}_d \udt{\omega}{\mu}{cb} \partial^c \gamma^{(1)b}_\beta \nonumber \\
& + 2 \gamma^{(0)\beta}_d \partial^\alpha \gamma^{(0)b}_\beta \udt{\omega}{\mu}{b\alpha} 
+ \gamma^{(
0)\beta}_d \gamma^{(1)b}_\beta \partial_\alpha \udut{\omega}{\mu}{b}{\alpha} \nonumber \\
&- \gamma^{(0)\beta}_d \partial_\nu \gamma^{(1)c}_\beta \udut{\omega}{\nu}{c}{\mu} - 
\eta^{\mu\alpha} \udt{\omega}{\beta}{db} \partial_\beta \gamma^{(1)b}_\alpha \nonumber \\
&+ \eta^{\mu\alpha} \udt{\omega}{\beta}{bd} \partial_\beta \gamma^{(1)b}_\alpha = 0.
\end{align}
Finally, note that if we choose our frame of reference to correspond to zero spin 
connection, then the above equation 
reduces to  
\begin{align}
&\eta^{\mu\alpha} \eta_{df} \Box \gamma^{(1)f}_\alpha - \Box \gamma^{(1)\mu}_d = 0,
\end{align}
where we have used the fact that 
\begin{equation}
\gamma^{(1)\nu}_b = - \gamma^{(0)\nu}_a \gamma^{(1)a}_\mu \gamma^{(0)\mu}_b, 
\end{equation}
which arises from   \eqref{eq:identity_tetrad-condition}.

In general, as it was mentioned in \cite{Krssak:2015oua}, in the case of non-zero spin 
connection it is hard even to extract the background solutions. Hence we can see 
that obtaining  the perturbed solution seems very difficult, since the 
background tetrad affects the perturbed solution. The detailed examination of the 
perturbed solutions in the case of $f(T)$ gravity with  non-zero spin 
connection lies beyond the 
scope of the present work.

We close this section by mentioning that the presented methodology can be 
extended to more general torsional modified gravitational theories, by defining 
appropriate gauge conditions on the tetrad, especially to theories 
in which the coordinate-indexed form of the field equations results in mixing between the 
metric and tetrad tensors. In this way, any information about the tetrad is not lost 
within the metric tensor, since the appropriate field equations are solved.

\subsection{Higher Order Metric Perturbations}

In this subsection we proceed to the analysis of higher-order perturbations, in order to 
understand  more transparently the potential deviations from general relativity. 
The standard approach is to consider perturbations around a flat Minkowski background. 
This is 
achieved by perturbing the metric tensor in the following manner
\begin{equation}
g_{\mu\nu} = \eta_{\mu\nu} + h^{(1)}_{\mu\nu} + h^{(2)}_{\mu\nu} + h^{(3)}_{\mu\nu} + 
\mathcal{O}\left(h^{(4)}_{\mu\nu}\right),
\end{equation}
where $|h^{(i)}_{\mu\nu}| << 1$, which is retained up to third order in this instance. 
Since the 
fundamental variable in the torsional formulation is the tetrad, the above metric 
perturbation can be obtained by 
the tetrad perturbation 
\begin{equation}
\udt{e}{a}{\mu} = \delta^a_\mu + \gamma^{(1)a}_\mu + \gamma^{(2)a}_\mu + 
\gamma^{(3)a}_\mu + \mathcal{O}\left(\gamma^{(4)a}_\mu\right),
\end{equation}
where $|\gamma^{(i)a}_\mu| << 1$. We remark that, in general, the zeroth order part of the 
tetrad 
perturbation is 
determined 
by the background metric, which in linearized gravity is usually the Minkowski metric. 
Thus, in the current case the zeroth order contribution to the tetrad perturbation 
turns out to be represented by the identity matrix, and that is why we introduced   the 
Kronecker delta. However, we mention that  this does not affect the obtained results, 
and the same conclusion is reached for other backgrounds, too.  

By the definition of the metric tensor $g_{\mu\nu} = \eta_{ab} \udt{e}{a}{\mu} 
\udt{e}{b}{\nu}$, we 
can relate the metric and tetrad perturbations through
\begin{align}
h^{(1)}_{\mu\nu} &= \eta_{ab} \left(\delta^a_\mu \gamma^{(1)b}_\nu + \gamma^{(1)a}_\mu 
\delta^b_\nu 
\right), \\
h^{(2)}_{\mu\nu} &= \eta_{ab} \left(\delta^a_\mu \gamma^{(2)b}_\nu + 
\gamma^{(1)a}_\mu\gamma^{(1)b}_
\nu + \gamma^{(2)a}_\mu \delta^b_\nu \right), \\
h^{(3)}_{\mu\nu} &= \eta_{ab} \Big(\delta^a_\mu \gamma^{(3)b}_\nu + 
\gamma^{(1)a}_\mu\gamma^{(2)b}_\nu + \gamma^{(2)a}_\mu\gamma^{(1)b}_\nu  + 
\gamma^{(3)a}_\mu \delta^b_\nu \Big). 
\end{align}
Inserting these expressions into the definition of the torsion tensor 
(\ref{eq:torsiontensordef}), and assuming for the moment zero spin connection, we 
obtain 
\begin{equation}
\udt{T}{a}{\mu\nu} = \partial_\mu \gamma^{(1)a}_\nu - \partial_\nu \gamma^{(1)a}_\mu 
+ 
\mathcal{O}\left(\gamma^{(2)a}_\mu\right),
\end{equation}
from which we can see that  the torsion tensor is at least of first order, with the 
zeroth-order contribution equal to zero. Consequently, from the definitions of the 
contorsion and superpotential tensors, namely relations (\ref{eq:contorsiondef}) and 
 (\ref{eq:superpotentialdef}) respectively, we deduce that they are both also at least of
first order since their zeroth order contributions are zero. 
Thus, the torsion scalar $T$, which is quadratic in the torsion tensor, becomes a 
second-order quantity. Finally, in order to handle the   $f(T)$ term, for simplicity we 
assume 
that this function is Taylor expandable around  $T=0$, namely
\begin{equation}
f(T) = f(0) + f_T(0) T  + \dfrac{1}{2!}f_{TT}(0) T^2 
 + \cdots .
 \label{TaylorfT1}
\end{equation}

Let us proceed by perturbing the equations of motion. According to Eq. 
(\ref{eq:global-eq-motion}), in the case of $f(T)$ gravity the field equations become
\begin{equation}
-f_T G_{\mu\nu} + \dfrac{1}{2}g_{\mu\nu}\left(f_T T - f\right) + 2{{S_{\nu}}^\alpha}_\mu 
\partial_\alpha f_T = 0.
\label{eomsfT}
\end{equation}
We mention that since we are interested in examining the properties of the gravitational 
waves, for simplicity we have neglected the contribution of the matter 
stress-energy tensor, namely  we neglect quadrupole moments which arise 
from the stress-energy tensor. 
 
Inserting the perturbed tetrad and metric in the field equations 
(\ref{eomsfT}), and under the Taylor expansion (\ref{TaylorfT1}), order by order we 
obtain 
\begin{align}
& \eta_{\mu\nu}f(0) = 0, \\
& f_T(0) G^{(1)}_{\mu\nu} = 0, \\
& f_T(0) G^{(2)}_{\mu\nu} = 0, \\
& f_T(0) G^{(3)}_{\mu\nu} +f_{TT}(0) T^{(2)} G^{(1)}_{\mu\nu}  
  - 2f_{TT}(0){S^{(1)\alpha}_{\nu}}_\mu \partial_\alpha T^{(2)} = 0.
\end{align}
Considering only the non-trivial case $f_T(0) \neq 0$ (otherwise GR cannot be obtained at 
any limit) the perturbed field equations 
simplify further to
\begin{align}
\label{zerthorderfT1}
& \eta_{\mu\nu}f(0) = 0, \\
& G^{(1)}_{\mu\nu} = 0, \\
& G^{(2)}_{\mu\nu} = 0, \\
& G^{(3)}_{\mu\nu} = 2\dfrac{f_{TT}(0)}{f_T(0)}{S^{(1)\alpha}_{\nu}}_\mu \partial_\alpha 
T^{(2)}.
\label{zerthorderfT3}
\end{align}

As we observe, the zeroth-order equation (\ref{zerthorderfT1}) implies that no 
cosmological constant is 
present  in the analysis, which was expected since the considered perturbations  are 
around a Minkowski background and not around a cosmological constant one. The first and 
second order equations coincide 
with the standard GR perturbed equations in vacuum. However, the new information is that  
at the third order equation (\ref{zerthorderfT3}) we find a 
deviation from the standard GR perturbation equation, with a contribution arising from 
the $f_{TT}$ term. Thus, the $f(T)$ effect on the perturbation equations enters 
only at the higher than second order, and the reason behind this is that the torsion 
scalar is quadratic in the torsion tensor. This is a radical difference with the 
case of curvature-based modified gravity, where the effect of the modification becomes 
manifest from first-order perturbation already. These features will become more 
transparent in the next section, where we study the case of $f(T,B)$ gravity.
Finally, note that in the   GR 
limit, i.e at $f_{TT}(0) = 0$, we re-obtain the standard GR results.

In summary, as we showed, in order to see the effect of $f(T)$ gravity on the 
gravitational 
waves themselves, one should look at third-order perturbations (higher-order 
contributions in curvature gravity have been examined in literature, see e.g. 
\cite{Matarrese:1997ay,Arcos:2015uqa}). Note that this concerns the effect on the 
``internal'' 
properties of the gravitational waves, as for instance in their polarization modes, where 
it was known that no further polarization modes are present in $f(T)$ gravity at 
first-order perturbation levels \cite{Bamba:2013ooa,Abedi:2017jqx}. However, we stress 
that in general the effect of $f(T)$ gravity on the cosmological gravitational wave 
propagation can be seen straightaway from the dispersion relation at first order, due to 
the effect of $f(T)$ gravity on the cosmological background itself \cite{Cai:2018rzd}.

\section{Gravitational Waves in $f(T,B)$ gravity}
\label{GwinfTB2}
 
In this section we will investigate the gravitational waves in the case of $f(T,B)$ 
gravity with action (\ref{eq:general-action}). From now on we consider only the 
case of zero spin connection, and we focus on the case $f(T,B)\neq f(T)$ since $f(T)$ 
gravity was investigated in the previous section. Furthermore, for convenience, we 
first study the 
gravitational waves around a Minkowski background, i.e in the case where a cosmological 
constant is absent from the  $f(T,B)$ form, and then we proceed to the general 
investigation of the case where a cosmological constant is allowed

\subsection{GWs in   $f(T,B)$ gravity in the absence of a cosmological constant}

We start with the perturbed metric around a Minkowski background:
\begin{equation}
g_{\mu\nu} = \eta_{\mu\nu} + h^{(1)}_{\mu\nu} + \mathcal{O}\left(h^{(2)}_{\mu\nu}\right),
\label{pertetradis000}
\end{equation}
where $|h^{(i)}_{\mu\nu}| << 1$. This metric perturbation can be obtained from the 
perturbed tetrad 
 
\begin{equation}
\udt{e}{a}{\mu} = \delta^a_\mu + \gamma^{(1)a}_\mu + 
\mathcal{O}\left(\gamma^{(2)a}_\mu\right).
\label{pertetradis}
\end{equation}
Using relation $g_{\mu\nu} = \eta_{ab} \udt{e}{a}{\mu} 
\udt{e}{b}{\nu}$ we 
acquire
\begin{align}\label{metric_pert222}
h^{(1)}_{\mu\nu} &= \eta_{ab} \left(\delta^a_\mu \gamma^{(1)b}_\nu + \gamma^{(1)a}_\mu 
\delta^b_\nu 
\right),
\end{align}
and thus for the perturbed torsion tensor we obtain 
\begin{equation}
\udt{T}{a}{\mu\nu} = \partial_\mu \gamma^{(1)a}_\nu - \partial_\nu \gamma^{(1)a}_\mu + 
\mathcal{O}\left(\gamma^{(2)a}_\mu\right).
\end{equation}

As we mentioned earlier,  the torsion tensor is at least first order, and thus
the torsion scalar $T$ is of second order in perturbations. This has a significant 
consequence, namely that relation (\ref{relation1}), specifically $R = -T+B$,  at first 
order
becomes  $R^{(1)} = B^{(1)}$ (we remind that $R$ is calculated using the 
Levi-Civita connection whilst $T$ and $B$ with the Weitzenb\"{o}ck connection).
Indeed, the Ricci scalar at first order is given to be
\begin{equation}
R^{(1)} = \eta^{\mu\nu}\partial_\rho\partial_\nu \udt{h}{(1)\rho}{\mu} - \Box h^{(1)},
\end{equation}
where indices are raised with respect to the Minkowski metric, $h^{(1)} \equiv 
\udt{h}{(1)\mu}{\mu}
$ and $\Box \equiv \partial_\mu \partial^\mu$. Expanding in terms of tetrads yields
\begin{equation}
R^{(1)} = 2 \delta_b^\rho \left(\eta^{\mu\nu} \partial_\nu \partial_\rho 
\gamma^{(1)b}_\mu - \Box \gamma^{(1)b}_\rho\right).
\end{equation}
On the hand, expanding the boundary term at first order yields
\begin{align}
B^{(1)} &= -2\left(\nabla^\mu \udt{T}{\nu}{\mu\nu}\right)^{(1)}  = 
-2\eta^{\mu\rho}\partial_\rho \udt{T}{(1)\nu}{\mu\nu} \nonumber\\
& = 2 \delta_b^\rho \left(\eta^{\mu\nu} \partial_\nu \partial_\rho \gamma^{(1)b}_\mu - 
\Box \gamma^{
(1)b}_\rho\right).
\end{align}
Thus, we can immediately see that at this order it is equal to the Ricci scalar.

In order to handle the $f(T,B)$ term for simplicity we assume 
that its form is Taylor expandable around the current values $T_0$ and $B_0$, namely
\begin{align}
f(T,B) =& f(T_0, B_0) + f_T(T_0,B_0) (T-T_0) \nonumber\\
& + f_B(T_0,B_0) (B-B_0) \nonumber\\
& + \dfrac{1}{2!}f_{TT}(T_0,B_0) (T-T_0)^2 \nonumber\\
& + \dfrac{1}{2!}f_{BB}(T_0,B_0) (B-B_0)^2 \nonumber\\
& + f_{TB}(T_0,B_0) (T-T_0)(B-B_0) + \dots \nonumber\\
& 
\end{align}
Furthermore, since we are only examining the properties of the gravitational waves, we 
neglect the matter sector.

Inserting all the above into the field equations of $f(T,B)$ gravity, namely 
Eq. (\ref{eq:global-eq-motion}), order by order we obtain
\begin{eqnarray}
&&
\!\!\!\!\!\!\!
\eta_{\mu\nu}f(0,0) = 0, \\
&&\!\!\!\!\!\!\!
-f_T(0,0) G^{(1)}_{\mu\nu}   + f_{BB}(0,0) \left(\eta_{\mu\nu} \Box - 
\partial_\mu 
\partial_\nu\right) R^{(1)} = 0,\ \ \  \
\label{eq:first-order-perturbation-TB}
\end{eqnarray}
where we have used the fact that $R^{(1)} = B^{(1)}$, and that $f(0,0) = 0$ from the 
zeroth order 
condition. The latter condition  is another statement for the fact that the arbitrary 
Lagrangian function does not include a cosmological constant. 

We proceed following \cite{Abedi:2017jqx} and  we define an effective mass by considering 
the trace of the first-order equation. This is also similar to the $f(R)$ gravity 
case. However, 
our effective mass is different to that of Ref. \cite{Abedi:2017jqx}. Indeed, by taking 
the trace
\begin{equation}\label{eq:trace-TB}
f_T(0,0) R^{(1)} + 3 f_{BB}(0,0) \Box R^{(1)} = 0, 
\end{equation}
we identify the effective mass $m$ by bringing the equation in the form $\left(\Box - 
m^2\right) R^{
(1)} = 0$, which turns out to be
\begin{equation}
m^2 \equiv -\dfrac{f_T(0,0)}{3 f_{BB}(0,0)}.
\end{equation}
We remark that in the $|m^2| \rightarrow \infty$ limit (for instance when $f_{BB}(0,0) = 
0$ and $f_T(0,0) \neq 0$), the equation reduces to that of GR. Since it is known that 
$f(T)$ gravity yields no further gravitational wave modes \cite{Bamba:2013ooa}, as we 
verified in the previous section, this special condition leads to a broader class of 
theories in which at first order yield the gravitational wave solutions.

In the case where $f_{BB}(0,0)\neq 0$ we can follow the procedure of $f(R)$ gravity 
\cite{Liang:2017ahj,Yang:2011cp,Capozziello:2008rq,Capozziello:2009nq,Berry:2011pb}  
(note that $f(R)$ is a particular subclass of $f(T,B)$ gravity, namely $f(-T+B)$ 
gravity). 
Firstly, we introduce the tensor $\bar{h}^{(1)}_{\mu\nu}$ to be
\begin{equation}\label{eq:trace-reversed}
h^{(1)}_{\mu\nu} = \bar{h}^{(1)}_{\mu\nu} - \dfrac{1}{2}\bar{h}^{(1)}\eta_{\mu\nu} + 
\dfrac{f_{BB}(
0,0)}{f_T(0,0)}\eta_{\mu\nu}R^{(1)},
\end{equation}
where $\bar{h}^{(1)}$ represents the trace of $\bar{h}^{(1)}_{\mu\nu}$. Similarly to 
the previous section we consider the non-trivial case  of
$f_T(0,0) \neq 0$ (otherwise GR cannot be obtained at any limit).
This 
simplifies Eq. \eqref{eq:first-order-perturbation-TB} to
\begin{equation}
\partial^\rho \partial_\nu \bar{h}^{(1)}_{\rho\mu} + \partial^\rho\partial_\mu 
\bar{h}^{(1)}_{\nu\rho} -\eta_{\mu\nu} \partial^\rho \partial^\alpha 
\bar{h}^{(1)}_{\rho\alpha} - \Box \bar{h}^{(1)}_{\mu\nu} = 0.
\end{equation}
As shown in \cite{Liang:2017ahj}, it is possible to consider the Lorenz gauge 
condition $\partial^\mu \bar{h}^{(1)}_{\mu\nu} = 0$, which simplifies the wave equation 
to 
\begin{equation}
\Box \bar{h}^{(1)}_{\mu\nu} = 0,
\end{equation}
as well as the traceless condition $\bar{h}^{(1)} = 0$. This allows for the solution
\begin{equation}
\bar{h}^{(1)}_{\mu\nu} = A_{\mu\nu} \exp\left(i k_\rho x^\rho\right),
\end{equation}
where $k_\rho$ is the four-wavevector, $A_{\mu\nu}$ are constant coefficients, $k_\rho 
k^\rho = 0$, 
$k^\mu A_{\mu\nu} = 0$ and $\udt{A}{\mu}{\mu} = 0$. The last conditions are the Lorenz 
gauge and 
traceless conditions respectively. On the other hand, the solution for 
 \eqref{eq:trace-TB} is
\begin{equation}\label{eq:Ricci-sol}
R^{(1)} = F \exp\left(i p_{\mu} x^{\mu}\right),
\end{equation}
where $F$ is a constant and $p_\mu$ is another four-wavevector such that $p_\mu p^\mu = 
-m^2$. 
Hence, the full solution for $h^{(1)}_{\mu\nu}$ is constructed as
\begin{align}
h^{(1)}_{\mu\nu} &=  A_{\mu\nu} \exp\left(i k_\rho x^\rho\right)  + 
\dfrac{f_{BB}(0,0)}{f_T(0,0)}\eta_{\mu\nu} F \exp\left(i p_{\mu} x^{\mu}\right).
\end{align}
Note that   from 
 \eqref{eq:first-order-perturbation-TB},\eqref{eq:trace-TB},
the Ricci tensor is found to be
\begin{equation}
R_{\mu\nu}^{(1)} = \dfrac{1}{6} \eta_{\mu\nu} R^{(1)} - \dfrac{f_{BB}(0,0)}{f_T(0,0)} 
\partial_\mu \partial_\nu R^{(1)}, 
\end{equation}
from which the solution of the Ricci scalar  \eqref{eq:Ricci-sol} simplifies to 
\begin{equation}
R_{\mu\nu}^{(1)} = \left(\dfrac{1}{6} \eta_{\mu\nu} -\dfrac{1}{3m^2} p_\mu p_\nu \right) 
R^{(1)}.
\end{equation}
Hence, it is trivial to verify that taking the trace yields a consistent relation for the 
Ricci scalar, as expected.

We proceed by analyzing the polarization states of the gravitational waves. As usual
we consider the geodesic deviation as in Ref. \cite{Liang:2017ahj}. We remark that 
although in teleparallel theories the particle motion is not described in terms of 
geodesics, mathematically one may still use the geodesic 
deviation formula, having in mind that all curvature quantities should obviously be 
calculated using the Levi-Civita connection \cite{aldrovandi2012teleparallel}  (for 
instance  see \cite{Darabi:2014dla} for the geodesic deviation in $f(T)$ gravity). 
Hence, we start from the geodesic 
deviation formula \cite{carroll2004spacetime}
\begin{equation}
\ddot{x}_i = -R_{i0j0} x^j,
\end{equation}
where dots represent coordinate time derivatives, $R_{\mu\nu\lambda\rho}$ is the Riemann 
tensor 
calculated with the Levi-Civita connection, $(t,x,y,z) = (0,1,2,3)$, $i = \lbrace 
1,2,3 \rbrace$ 
and $x^j = (x, y, z)$. Moreover, we consider the signature $(+,-,-,-)$, and for 
simplicity we assume that the wave   propagates in the $z$-direction.  

From the perturbation analysis presented above, we find that 
 \begin{equation}
R_{i0j0} = \dfrac{1}{2}k_0^2 \bar{h}^{(1)}_{ij} - \dfrac{1}{6m^2}\left[\eta_{ij} p^2_0 
R^{(1)} + p_
i p_j R^{(1)}\right].
\end{equation}
Therefore, the geodesic deviation becomes
\begin{align}
\ddot{x} &= \left[\dfrac{1}{2}k_0^2 \bar{h}^{(1)}_{+} + \dfrac{1}{6m^2} p^2_0 
R^{(1)}\right] x + \dfrac{1}{2}k_0^2 \bar{h}^{(1)}_{\times} y, \\
\ddot{y} &= \left[-\dfrac{1}{2}k_0^2 \bar{h}^{(1)}_{+} + \dfrac{1}{6m^2} p^2_0 
R^{(1)}\right] y + \dfrac{1}{2}k_0^2 \bar{h}^{(1)}_{\times} x, \\
\ddot{z} &= \dfrac{1}{6m^2}\left(p^2_0 - p^2_3 \right) R^{(1)} z = -\dfrac{1}{6} R^{(1)} 
z,
\end{align}
where in the last equation we have used that $p_\mu p^\mu = -m^2$. Additionally, since 
the wave propagates in the $z$-direction, we have used and defined $\bar{h}^{(1)}_{11} = 
-\bar{h}^{(1)}_{22} \equiv \bar{h}
^{(1)}_{+}$ and $\bar{h}^{(1)}_{12} = \bar{h}^{(1)}_{21} \equiv \bar{h}^{(1)}_{\times}$, 
which 
represent the massless $+$ and $\times$ polarizations.

As we observe, in the TEGR limit, namely at $|m^2| 
\rightarrow \infty$ and $R^{(1)} \rightarrow 0$, the remaining modes are the $+$ 
and $\times$ polarizations  as expected. However, in the case $|m^2| < \infty$ we find 
the presence of the longitudinal and breathing modes in the geodesic deviation equations. 
This is one of the main results of the present work, namely that $f(T,B)$ gravity, in 
the case where  $f(T,B)\neq f(T)$, does 
have further polarization modes at first-order perturbation, in contrast to the case of 
$f(T)$ gravity. The reason for this behavior is the fact that although the first-order 
perturbation does not have any effect on $T$, it does affect the boundary term $B$.

\subsection{Tetrad Solutions for GWs in $f(T,B)$ gravity} 

In the previous subsection we analyzed the gravitational waves in $f(T,B)$ gravity from 
the metric perturbation side. We now proceed  to their examination from the tetrad 
perturbation side. In order to do this we start from the perturbed tetrad 
(\ref{pertetradis}) and we insert it into the tetrad form of the  $f(T,B)$ field 
equations, namely into Eq. \eqref{eq:general-field-equations}. Neglecting the matter 
sector, order by order we obtain
\begin{align}
&\delta^\mu_a f^{(0)} = 0, \\
&\delta^\mu_a \Box f_B^{(1)} - \delta_a^\nu \partial^\mu \partial_\nu f_B^{(1)} + 
\dfrac{1}{2}B^{(1)
} f_B^{(0)} \delta^\mu_a \nonumber\\
&\ \ + 2\partial_\nu \dut{S}{a}{(1)\nu\mu} f_T^{(0)} - \dfrac{1}{2}\delta_a^\mu f^{(1)} = 
0.
\label{perteq1gwtetrad}
\end{align}
As before, the zeroth-order condition is a verification that there is no cosmological 
constant present, i.e. that the perturbation is performed around Minkowski background. In 
order to simplify the first-order equation we remark that $f^{(1)} = f_B^{(0)} B^{(1)}$ 
and $f_B^{(1)} = f_{BB}^{(0)} B^{
(1)}$, and as usual we consider the non-trivial case   $f_T^{(0)}\neq0$.
 Therefore, Eq. (\ref{perteq1gwtetrad}) reduces to
\begin{equation}
\partial_\nu \dut{S}{a}{(1)\nu\mu} + \dfrac{f_{BB}^{(0)}}{2f_T^{(0)}} \left[\delta^\mu_a 
\Box B^{(1)
} - \delta_a^\nu \partial^\mu \partial_\nu B^{(1)}\right] = 0.
\end{equation}

Due to the introduction of the $B^{(1)}$ terms in the above equation, the traceless and 
Lorenz conditions used for the simple case of $f(T)$ gravity in subsection 
\ref{Tetrad_Solutions_for_GWs} need to be modified to accommodate 
a more suitable gauge choice. From the metric 
approach of the previous subsection we instead have the ``trace-reversed'' metric 
$\bar{h}_{\mu\nu}$ in  \eqref{eq:trace-reversed}, which satisfies the traceless and 
Lorenz gauge conditions $\bar{h} = 0$, and $  \partial^\mu \bar{h}_{\mu\nu} = 0$. 
Therefore, the ``traceless'' condition becomes
\begin{equation}
\delta^\nu_b \gamma^{(1)b}_\nu = \dfrac{2f_{BB}^{(0)} B^{(1)}}{f_T^{(0)}},
\end{equation}
whilst the ``Lorenz condition'' reads as 
\begin{equation}
\eta_{ab} \left(\partial^a \gamma^{(1)b}_\nu + \delta^b_\nu \partial^\mu 
\gamma^{(1)a}_\mu\right) = 
\dfrac{f_{BB}^{(0)} \partial_\nu B^{(1)}}{f_T^{(0)}}.
\end{equation}
In this way, the field equations simplify to
\begin{equation}
A^\mu_a \equiv   \eta^{\mu\alpha} \eta_{ab} \Box \gamma^{(1)b}_\alpha + \delta^\mu_b 
\delta^\rho_
a \Box  \gamma^{(1)b}_\rho 
 - \delta^\mu_a \dfrac{f_{BB}^{(0)} \Box B^{(1)}}{f_T^{(0)}} = 0,
 \label{firstordereq5}
\end{equation}
which yields the following system of field equations
\begin{align}
\!
&A^0_0: & & \Box \left(\gamma^{(1)0}_0 - \dfrac{f_{BB}^{(0)} B^{(1)}}{2f_T^{(0)}}\right) 
= 0, \\ \!
&A^0_i = -A_0^i: & &\Box \left(\gamma^{(1)0}_i - \gamma^{(1)i}_0\right) = 0, 
\end{align}
\begin{align}
&A^i_j \left(i \neq j\right): & &\Box \left(\gamma^{(1)j}_i + \gamma^{(1)i}_j\right) = 0, 
\\
&A^i_m \left(i = m\right): & &\Box \left(\gamma^{(1)i}_i - \dfrac{f_{BB}^{(0)} 
B^{(1)}}{2f_T^{(0)}}\right) = 0,
\end{align}
where $\eta_{\mu\nu} = \text{diag}\left(-1,1,1,1\right)$ and 
$i, j = \{1,2,3\}$. Moreover,  since we are working in the Minkowski metric Cartesian 
coordinate system, the indices $\{0,
1,2,3\}$ correspond to $\{t,x,y,z\}$ respectively. 

The above equations are standard wave equations, and thus we assume a 
plane-wave solution by working in Fourier space. Without loss of generality, we shall 
assume that the waves propagate  in the $z$-direction. Hence, the solution for the 
perturbed tetrad is 
\begin{widetext}
\begin{equation}
\resizebox{\linewidth}{!}
{%
$\gamma^{(1)a}_\mu = \begin{pmatrix}
 A\exp (ik_\mu x^\mu) + \dfrac{f_{BB}^{(0)} B^{(1)}}{2f_T^{(0)}} & \gamma^{(1)1}_0 & 
\gamma^{(1)2}_
0 & \gamma^{(1)3}_0 \\
B_1\exp (ik_\mu x^\mu) + \gamma^{(1)1}_0 & D\exp (ik_\mu x^\mu) + \dfrac{f_{BB}^{(0)} 
B^{(1)}}{2f_
T^{(0)}} & \gamma^{(1)2}_1 & \gamma^{(1)3}_1 \\
B_2\exp (ik_\mu x^\mu) + \gamma^{(1)2}_0 & C \exp (ik_\mu x^\mu)-\gamma^{(1)2}_1 & -D\exp 
(ik_\mu 
x^\mu) + \dfrac{f_{BB}^{(0)} B^{(1)}}{2f_T^{(0)}} & \gamma^{(1)3}_2 \\
\gamma^{(1)3}_0 - 2A\exp (ik_\mu x^\mu) & B_1\exp (ik_\mu x^\mu)-\gamma^{(1)3}_1 & 
B_2\exp (ik_\mu 
x^\mu)-\gamma^{(1)3}_2 & -A\exp (ik_\mu x^\mu) + \dfrac{f_{BB}^{(0)} B^{(1)}}{2f_T^{(0)}}
\end{pmatrix}.  
\label{tetradsol2basic}$
}
\end{equation}
\end{widetext}
Here, the $\gamma^{(1)j}_i$ are undetermined tetrad components, not constrained by the 
equations, $A, B_{1,2}, C$ and $D$ are constants, and $k_\mu$ is the wave-vector such 
that 
$k_\mu k^\mu = 0$. 

As we observe, in the $f_{BB}^{(0)} \rightarrow 0$ limit the first-order perturbed 
equation (\ref{firstordereq5}) reduces to that of GR and $f(T)$ gravity, and hence the 
tetrad solution should describe the same solution. Comparing with the tetrad solution 
obtained in the case of simple $f(T)$ gravity, namely solution (\ref{tetradsolbasic1}),
we deduce that this is obtained by setting 
$A = B_{1,2} = 0$, implying that these constants reflect the transverse property of the 
$h_+ \equiv 2D\exp (ik_\mu x^\mu)$ and $h_\times \equiv C \exp (ik_\mu x^\mu)$ 
polarizations and hence can be 
removed. This can also be identified from the   metric tensor solution corresponding to 
(\ref{tetradsol2basic}), namely
\begin{widetext}
\begin{equation}
h^{(1)}_{\mu\nu} = 
\begin{pmatrix}
-2A\exp (ik_\mu x^\mu) - \dfrac{f_{BB}^{(0)} B^{(1)}}{f_T^{(0)}} & B_1\exp (ik_\mu x^\mu) 
& B_2\exp 
(ik_\mu x^\mu) & -2A\exp (ik_\mu x^\mu) \\
B_1\exp (ik_\mu x^\mu) & h_+ + \dfrac{f_{BB}^{(0)} B^{(1)}}{f_T^{(0)}}  & h_\times & 
B_1\exp (ik_\mu x^\mu) \\
B_2\exp (ik_\mu x^\mu) & h_\times & -h_+ + \dfrac{f_{BB}^{(0)} B^{(1)}}{f_T^{(0)}} & 
B_2\exp (ik_\mu x^\mu) \\
-2A\exp (ik_\mu x^\mu) &  B_1\exp (ik_\mu x^\mu) &  B_2\exp (ik_\mu x^\mu) & -2A\exp 
(ik_\mu x^\mu) 
+ \dfrac{f_{BB}^{(0)} B^{(1)}}{f_T^{(0)}}
\end{pmatrix},
\end{equation}
\end{widetext}
which in the $A, B_{1,2}, f_{BB}^{(0)} \rightarrow 0$ limit reduces to the standard 
perturbed 
metric solution for waves traveling in the $z$-direction.

\subsection{GWs in   $f(T,B)$ gravity in the presence of a cosmological constant}

In the previous subsection we investigated the gravitational waves in $f(T,B)$ gravity 
through a perturbation around a Minkowski background in vacuum, i.e. in the absence of a 
cosmological constant in the form of  $f(T,B)$. In the present subsection we examine the 
contribution of a cosmological constant to the gravitational waves following the 
procedure of \cite{Naf:2008sf,Bernabeu:2011if}.
%with a modification to the tetrad 
%perturbations arising. 
Implications of the cosmological constant onto gravitational waves 
(for instance
quadropole contributions and effects during the cosmological history), in the case of   
other scenarios, have been investigated in  
\cite{Ashtekar:2017dlf,Bicak:1999ha,Bicak:1999hb,Abramo:1997hu,Capozziello:2007vd}.

Let $\Lambda$ denote the cosmological constant. Similarly to the previous analysis   
we shall again consider a linearized gravity approach, however, since the cosmological 
constant also affects the background metric, the perturbations will not be performed 
around a Minkowski background but around a background that incorporates  the 
contributions of $\Lambda$. This is achieved 
through
\begin{equation}
g_{\mu\nu} = h^{(0)}_{\mu\nu} + h^{(1)}_{\mu\nu} + 
\mathcal{O}\left(h^{(2)}_{\mu\nu}\right),
\end{equation}
where $|h^{(i)}_{\mu\nu}| << 1$, and
\begin{align}
h^{(0)}_{\mu\nu} &= \eta_{\mu\nu} + \Lambda h^{(0)\Lambda}_{\mu\nu} + 
\mathcal{O}\left(\Lambda^2\right), \\
h^{(1)}_{\mu\nu} &= h^{(1)\text{GW}}_{\mu\nu} + \Lambda h^{(1)\text{GW}\Lambda}_{\mu\nu} 
+ \mathcal{
O}\left(\Lambda^2\right),
\end{align}
where $h^{(0)\Lambda}_{\mu\nu}$ refers to the first-order contribution of $\Lambda$ to 
the background metric, $h^{(1)\text{GW}}_{\mu\nu}$ is the gravitational wave perturbation 
without the effect of $\Lambda$, and $h^{(1)\text{GW}\Lambda}_{\mu\nu}$ is the 
contribution of the cosmological constant to the gravitational wave perturbation. 
Correspondingly, the tetrad perturbation can be 
constructed as 
\begin{equation}
\udt{e}{a}{\mu} = \gamma^{(0)a}_\mu + \gamma^{(1)a}_\mu + 
\mathcal{O}\left(\gamma^{(2)a}_\mu\right),
\end{equation}
where $|\gamma^{(i)a}_\mu| << 1$, and 
\begin{align}
\gamma^{(0)a}_\mu &= \delta^a_\mu + \Lambda \gamma^{(0,\Lambda)a}_\mu + 
\mathcal{O}(\Lambda^2), \\
\gamma^{(1)a}_\mu &= \gamma^{(1,\text{GW})a}_\mu + \Lambda 
\gamma^{(1,\text{GW}\Lambda)a}_\mu + \mathcal{O}(\Lambda^2),
\end{align}
where $\gamma^{(0,\Lambda)a}_\mu$ refers to the first-order contribution of $\Lambda$ to 
the background tetrad, $\gamma^{(1,\text{GW})a}_\mu $ is the gravitational wave 
perturbation without the effect of $\Lambda$, and $\gamma^{(1,\text{GW}\Lambda)a}_\mu$ is 
the contribution of the cosmological constant to the gravitational wave perturbation. 
Therefore, the metric perturbations are related to the tetrad perturbations through
\begin{align}
h^{(0)\Lambda}_{\mu\nu} &= \eta_{ab} \left(\delta^a_\mu \gamma^{(0,\Lambda)b}_\nu + 
\gamma^{(0,\Lambda)a}_\mu \delta^b_\nu\right), \\
h^{(1)\text{GW}}_{\mu\nu} &= \eta_{ab} \left(\delta^a_\mu\gamma^{(1,\text{GW})b}_\nu + 
\gamma^{(1,\text{GW})a}_\nu \delta^b_\mu \right), \\
h^{(1)\text{GW}\Lambda}_{\mu\nu} &= \eta_{ab} \Big(\delta^a_\mu 
\gamma^{(1,\text{GW}\Lambda)b}_\nu 
+ 2 \gamma^{(1,\text{GW})a}_\mu \gamma^{(0,\Lambda)b}_\nu \nonumber\\
& \ \ \ + \gamma^{(1,\text{GW}\Lambda)a}_\mu \delta^b_\nu\Big).
\end{align}
To facilitate the perturbation analysis, we again assume that the function $f(T,B)$ is 
Taylor 
expandable about $T = B = 0$, namely
\begin{align}
f(T,B) &= f(0,0) + f_T(0,0) T + f_B(0,0) B \nonumber \\
& + \dfrac{1}{2!}f_{TT}(0,0) T^2+ \dfrac{1}{2!}f_{BB}(0,0) B^2 \nonumber \\
& + f_{TB}(0,0) TB + \cdots\, .
\end{align}
Hence, we can identify $f(0,0) \equiv 2\Lambda$, since this term  behaves as a 
cosmological constant, which can 
be seen when it is substituted in the equations of motion or in the action. The choice of 
this value is in order to ease the comparisons with the cosmological constant present in 
GR.

Inserting the above into the field equations  
\eqref{eq:global-eq-motion} and neglecting the matter sector, for the 
zeroth-order perturbation equation we obtain
\begin{equation}
-f_T^{(0)} G^{(0)}_{\mu\nu} + \dfrac{1}{2}h^{(0)}_{\mu\nu} \Big(f_B^{(0)} B^{(0)} 
 + f_T^{(0)} T^{(0)} - f^{(0)}\Big) = 0,
\end{equation}
which when expanded over $\Lambda$ order yields
\begin{equation}
f_T(0,0) G^{(0)\Lambda}_{\mu\nu} + \dfrac{1}{2}\eta_{\mu\nu}f(0,0) = 0,
\end{equation}
where the superscript $\Lambda$ refers to the cosmological constant contribution of the 
function (from here onwards, this shall be assumed for all symbols). Taking the trace 
yields 
\begin{equation}
f_T(0,0) R^{(0)\Lambda} = 2f(0,0).
\end{equation}
In the case of TEGR with cosmological constant, i.e. for $f(T,B) = T+2\Lambda$, the above 
expression yields precisely the standard result for GR with a cosmological constant, 
namely $R = 4\Lambda$. Note that the 
first derivative $f_T(0,0)$ rescales this value of the Ricci scalar.  

Additionally, the first-order perturbation  becomes
\begin{align}
&-f^{(0)}_T G^{(1)}_{\mu\nu} -f^{(1)}_T G^{(0)}_{\mu\nu} + \left[h^{(0)}_{\mu\nu} 
\Box^{(0)} - \nabla^{(0)}_\mu \nabla^{(0)}_\nu\right] f^{(1)}_B \nonumber \\
& + \dfrac{1}{2}h^{(0)}_{\mu\nu}\left[f^{(0)}_B B^{(1)} + f^{(1)}_B B^{(0)}+ f^{(0)}_T 
T^{(1)} + f^{
(1)}_T T^{(0)} - f^{(1)}\right] \nonumber \\
&
+ \dfrac{1}{2}h^{(1)}_{\mu\nu}\left[f^{(0)}_B B^{(0)} + 
f^{(0)}_T 
T^{(0)} - f^{(0)}\right] \nonumber \\
& + 2{{S^{(0)\alpha}_{\nu}}}_\mu \partial_\alpha \left(f^{(1)}_T + f^{(1)}_B\right) + 
2{{S^{(1)\alpha}_{\nu}}}_\mu \partial_\alpha \left(f^{(0)}_T + f^{(0)}_B\right) = 0.
\end{align}
Expanding over $\Lambda$ order yields
\begin{align}
&f_T(0,0)-G^{(1)\text{GW}} \nonumber \\
& + f_{BB}(0,0) \left(\eta_{\mu\nu} 
\Box^{(0)\text{BG}} - 
\partial_\mu \partial_\nu\right) B^{(1)\text{GW}} = 0, 
\label{firsreqaution1}
\end{align}
and 
\begin{align}
&-f_T(0,0)G^{(1)\text{GW}\Lambda}-G^{(1)\text{GW}}f_{TB}(0,0)B^{(0)\Lambda} \nonumber \\
&
-f_{TB}(0,0)B^{(1)\text{
GW}} G^{(0)\Lambda}_{\mu\nu} \nonumber \\
&+f_{BB}(0,0) \left[h^{(0)\Lambda}_{\mu\nu} \Box^{(0)\text{BG}} + \eta_{\mu\nu} 
\Box^{(0)\Lambda} - 
\nabla^{(0)\Lambda}_\mu \partial_\nu\right] B^{(1)\text{GW}} \nonumber \\
&+ \!\left(\!\eta_{\mu\nu} \Box^{(0)} \!-\! \partial_\mu \partial_\nu\right) \!\!
\left[\!f_{BB}(0,\!0)B^{(1)\text{
GW}\Lambda}\! + \!f_{TB}(0,\!0)T^{(1)\text{GW}\Lambda}\!\right]\nonumber \\
& + 
\dfrac{1}{2}\eta_{\mu\nu}f_{BB}(0,0)B^{(0)
\Lambda}B^{(1)\text{GW}}  - \dfrac{1}{2}h^{(1)\text{GW}}_{\mu\nu}f(0,0) \nonumber \\
&+ 2\left[f_{TB}(0,0)+ 
f_{BB}(0,0)\right]{{S^{(0,\Lambda)\alpha}_{\nu}}}_\mu \partial_\alpha B^{(1)\text{GW}} 
\nonumber \\
&+ 2\left[f_{TB}(0,0)+ f_{BB}(0,0)\right]{{S^{(1,\text{GW})\alpha}_{\nu}}}_\mu 
\partial_\alpha B^{(
0)\Lambda} = 0,
\label{secondreqaution1}
\end{align} 
where the superscript BG refers to the Minkowski metric contribution. It is clear that 
Eq. (\ref{firsreqaution1}) is exactly the same with 
(\ref{eq:first-order-perturbation-TB}) (recalling that $R^{(1)} = B^{(1)}$). On the 
other hand, Eq. (\ref{secondreqaution1}) describes the relation 
between the standard gravitational waves and the cosmological constant contribution to 
them, which is the main result of the present subsection.

Finally, it is interesting to note that for the case of simple $f(T)$ gravity, namely for 
$f(T,B) = f(T)$, equations (\ref{firsreqaution1}) and (\ref{secondreqaution1}) reduce to
\begin{align}
&f_T(0)G^{(1)\text{GW}} = 0, \\
&f_T(0)G^{(1)\text{GW}\Lambda} + \dfrac{1}{2}h^{(1)\text{GW}}_{\mu\nu}f(0) = 0.
\end{align}
Thus, for 
$f(T)$ gravity at first order, the effect of the cosmological constant on the 
gravitational waves is also affected by the value of $f_T(0)$, as was also found in 
\cite{Cai:2018rzd}. Lastly, in the case of TEGR these equations match identically to 
those 
of GR, as expected.

\section{Gravitational Waves  in  $f(T,T_G)$  gravity}
\label{GWinTTG}

In this section we proceed to the investigation of the gravitational waves in another 
class of modified teleparallel gravity, namely  $f(T,T_G)$  gravity, in which one uses in 
the Lagrangian the teleparallel equivalent of the Gauss-Bonnet combination $T_G$. In 
particular, in curvature-based modified gravity one may add in the Lagrangian 
functions of the higher-order Gauss-Bonnet invariant, defined as
\begin{equation}
G = R_{\mu\nu\lambda\sigma}R^{\mu\nu\lambda\sigma}-4R_{\mu\nu}R^{\mu\nu}+R^2,
\end{equation}
and construct $f(R,G)$ theories  \cite{Nojiri:2005jg,Bamba:2009uf}.
Correspondingly, one can construct its teleparallel equivalent $T_G$, which reads as 
 \cite{Kofinas:2014owa} 
\begin{align}
&
\!\!\!\!
T_G=  
\big(K^{\alpha}_{\;\;\gamma\beta}K^{\gamma\lambda}_{\;\;\;\;\rho}K^{\mu}_{
\;\;\epsilon\sigma}K^{\epsilon\nu}_{\;\;\;\;\varphi} 
-2K^{\alpha\lambda}_{\;\;\;\;\beta}K^{\mu}_{\;\;\gamma\rho} K^{\gamma}
_{\;\;\epsilon\sigma}K^{\epsilon\nu}_{\;\;\;\;\varphi} \nonumber\\
&
\!\!\!\!
+2K^{\alpha\lambda}_{\;\;\;\;\beta}K^{\mu}_{\;\;\gamma\rho}K^{\gamma\nu}_{
\;\;\;\;\epsilon}K^{\epsilon}_{\;\;\sigma\varphi}+  
2K^{\alpha\lambda}_{\;\;\;\;\beta}K^{\mu}_{\;\;\gamma\rho}K^{\gamma\nu}_{\;\;\;\;\sigma,
\varphi}\Big)\delta^{\beta\rho\sigma\varphi}_{\alpha\lambda\mu\nu},
\end{align}
and use it the Lagrangian, resulting to the  so-called $f(T,T_G)$ gravity. The field 
equations of 
$f(T,T_G)$ 
gravity write as \cite{Kofinas:2014owa,Kofinas:2014daa}
\begin{align}
&2\left(H^{[ac]b} + H^{[ba]c} - H^{[cb]a}\right)_{,c} \nonumber\\
& + 2\left(H^{[ac]b} + H^{[ba]c} - H^{[cb]a}\right)\udt{C}{d}{dc} \nonumber\\
& + \left(2H^{[ac]d} + H^{dca}\right)\udt{C}{b}{cd} \nonumber\\
& + 4H^{[db]c}\dut{C}{(dc)}{a} + \udt{T}{a}{cd}H^{cdb} - h^{ab} \nonumber\\
& + \left(f - Tf_T -T_G f_{T_G}\right)\eta^{ab} = 0,
\end{align}
where
\begin{widetext}
\begin{align}
&H^{abc} = f_T \left(\eta^{ac} \udt{K}{bd}{d} - K^{bca}\right) + 
\epsilon^{cprt}\udt{\epsilon}{a}{
kdf}\left[\left(f_{T_G}\udt{K}{bk}{p}\udt{K}{df}{r}\right)_{,t} + f_{T_G} \udt{C}{q}{pt} 
\udt{K}{bk}
{[q}\udt{K}{df}{r]}\right] \nonumber \\
&+ f_{T_G}\bigg[\epsilon^{cprt}\left(2\udt{\epsilon}{a}{dkf}\udt{K}{bk}{p}\udt{K}{d}{qr} 
+ \epsilon_
{qdkf}\udt{K}{ak}{p}\udt{K}{bd}{r} + 
\udt{\epsilon}{ab}{kf}\udt{K}{k}{dp}\udt{K}{d}{qr}\right)\udt{
K}{qf}{t} \nonumber \\
&+ \epsilon^{cprt} \udt{\epsilon}{ab}{kd} \udt{K}{fd}{p} \left(\udt{K}{k}{fr,t} - 
\dfrac{1}{2}\udt{
K}{k}{fq} \udt{C}{q}{tr}\right) + \epsilon^{cprt} \udt{\epsilon}{ak}{df} 
\udt{K}{df}{p}\left(\udt{K}
{b}{kr,t}-\dfrac{1}{2}\udt{K}{b}{kq}\udt{C}{q}{tr}\right)\bigg], \\
&\udt{C}{c}{ab} = \dut{E}{a}{\mu} \dut{E}{b}{\nu} \left(\udt{e}{c}{\mu,\nu} - 
\udt{e}{c}{\nu,\mu}\right), \\
&h^{ab} = f_T \udt{\epsilon}{a}{kcd}\epsilon^{bpqd} \udt{K}{k}{fp}\udt{K}{fc}{q}.
\end{align}
\end{widetext}
Note that for the purpose of the gravitation wave investigation, the stress-energy 
contribution has been neglected.

 In order to analyze the gravitational waves in this theory, we again consider the linear 
perturbations in the metric  and in the tetrad around a Minkowski background, 
namely expressions (\ref{pertetradis000}) and (\ref{pertetradis}) respectively. From 
the definition of $T_G$ we deduce that  is at least a fourth-order quantity in the 
tetrad perturbation. Furthermore, for 
simplicity we assume 
that $f(T,T_G)$ is Taylor expandable around $T = T_G = 0$. 
Therefore, the resulting zeroth and first-order perturbation equations are 
\begin{align}
&\eta^{ab}f(0,0) = 0, \\
&\left(H^{(1)[ac]b} + H^{(1)[ba]c} - H^{(1)[cb]a}\right)_{,c} = 0.
\label{firstordereqTG}
\end{align}
Similarly to the previous discussion, the zeroth-order equation implies that no 
cosmological constant is present in the theory, which is consistent to the fact that
the perturbations are performed around Minkowski background.

From the 
definition of $H^{abc}$ we remark that 
\begin{equation}
H^{(1)abc} = f_T(0,0) \left(\eta^{ac} \udt{K}{(1)bd}{d} - K^{(1)bca}\right).
\end{equation}
Substituting in the first-order equation (\ref{firstordereqTG}) yields
\begin{align}
f_T(0,0) &\bigg[\eta^{ab} \udt{K}{(1)cd}{d} - \eta^{ac} \udt{K}{(1)bd}{d} + K^{(1)bca} 
\bigg]_{,c} \nonumber\\
& = f_T(0,0) \partial_c S^{(1)abc} = 0,
\label{eqcase1TG}
\end{align}
where the definition of the superpotential (\ref{eq:superpotentialdef}) has been used. 
Interestingly enough, for the non-trivial case $ f_T(0,0)\neq0$ (otherwise GR cannot be 
obtained at any limit) Eq. (\ref{eqcase1TG}) coincides with  Eq. (\ref{eqcase1}) obtained 
in the case of simple $f(T)$ gravity (note that changing coordinate to tangent-space
indices involves only the zeroth-order part of the tetrad which is just the Kronecker 
delta). Hence, we deduce that at first-order perturbation level, the gravitational waves 
behave in the same way in $f(T)$ and $f(T,T_G)$ gravities, and thus the previously 
obtained result that no further polarization modes comparing to GR are obtained at this 
order holds for $f(T,T_G)$ gravity too. This result was expected, since $T_G$ is a 
higher-order torsion invariant and therefore its effect switches on at higher 
perturbation 
orders.

\section{Conclusions}
\label{Conclusion}

In this work we investigated the gravitational waves and their properties, in various 
modified teleparallel theories, such as $f(T), f(T,B)$ and $f(T,T_G)$ gravities, by 
utilizing  the perturbed equations. Furthermore, we performed the analysis in 
both the metric and tetrad language, in order to reveal the properties of the formalism. 
Additionally, we performed the perturbations around a Minkowski background, case which is 
obtained in the absence of a cosmological constant, but also in the case where the 
presence of a cosmological constant changes the background around which the perturbations 
are realized. Finally, in the case of usual $f(T)$ gravity we performed the analysis both 
for the standard formulation of zero spin connection, as well as for the most general and 
fully covariant case of a non-zero spin connection.

For the case of simple $f(T)$ gravity we verified the result that no further polarization 
modes comparing to GR are present at first-order perturbation level, since the torsion 
scalar (which is quadratic in the torsion tensor) does not acquire any perturbative 
contribution at this level. Hence, as we showed, in order to see the effect of $f(T)$ 
gravity on the gravitational waves themselves one should look at  third-order 
perturbations, in which a deviation from GR is obtained due the contribution from the 
$f_{TT}$ component. Nevertheless, we mention that this is the effect on the ``internal'' 
properties of the gravitational waves, such as   their polarization modes, since
in general the effect of $f(T)$ gravity on the cosmological gravitational wave 
propagation can be seen straightaway from the dispersion relation at first order, due to 
the effect of $f(T)$ gravity on the cosmological background on which the gravitational 
waves propagate \cite{Cai:2018rzd}.

For the case of $f(T,B)$ gravity with $f(T,B)\neq f(T)$, by examining the geodesic 
deviation equations, we showed that extra polarization modes, namely the longitudinal and 
breathing modes, do appear at first-order perturbation level. The reason for this 
behavior 
is the fact that although the first-order perturbation does not have any effect on $T$, 
it 
does affect the boundary term $B$. Additionally, in the case where a cosmological 
constant 
is present we have extracted the gravitational wave equations, obtaining the 
cosmological-constant corrections to the solutions, which reflect the fact that the 
background is not Minkowski anymore. 

Finally, we investigated the gravitational waves in $f(T,T_G)$ gravity, which at 
first-order perturbation level exhibit the same behavior to those of $f(T)$ gravity, that 
is they do not have extra polarization modes comparing to GR. This result was expected, 
since the teleparallel equivalent of the Gauss-Bonnet term $T_G$ is a higher-order 
torsion invariant and therefore its effect switches on at higher perturbation orders.

In summary, as we showed, apart from their difference from curvature-based gravity, 
different modified teleparallel gravities exhibit different gravitational wave properties 
amongst themselves, despite the fact that at first sight they might look similar 
theories. 
Hence, the advancing gravitational-wave astronomy would help to alleviate the degeneracy 
not only between curvature and torsional modified gravity, but also between different 
subclasses of modified teleparallel gravities.

\section*{Acknowledgments}
 
The research work disclosed in this paper is partially funded by the ENDEAVOUR 
Scholarships Scheme.
This article is based upon work from CANTATA COST (European Cooperation in Science and 
Technology) 
action CA15117, EU Framework Programme Horizon 2020.

\end{document}